\documentclass{ws-ijmpcs}

\usepackage{epsfig,multicol,multirow}
\usepackage{amsmath,subfigure,latexsym,amssymb,cite}
\usepackage{bbm}

\newcommand{\1}{\mbox{1}\hspace{-0.25em}\mbox{l}}
\newcommand {\SO}{\mathop{\rm SO}}

\newcommand {\beq} {\begin{equation}}
\newcommand {\eeq} {\end{equation}}
\newcommand {\beqa}{\begin{eqnarray}}
\newcommand {\eeqa}{\end{eqnarray}}

\newcommand {\tr}{{\rm tr\,}}

\begin{document}


\markboth{Jun Nishimura}
{(3+1)-Dimensional Expanding Universe from a Lorentzian Matrix Model...} 

%
\catchline{}{}{}{}{}
%

\title{(3+1)-DIMENSIONAL EXPANDING UNIVERSE FROM A LORENTZIAN MATRIX MODEL 
FOR SUPERSTRING THEORY IN (9+1)-DIMENSIONS}

\author{JUN NISHIMURA}

\address{KEK Theory Center, High Energy Accelerator Research Organization\\
1-1 Oho, Tsukuba, Ibaraki 305-0801, Japan\\
jnishi@post.kek.jp}



\maketitle


\begin{abstract}
We study the Lorentzian version of
the type IIB matrix model as a nonperturbative
formulation of superstring theory in (9+1)-dimensions.
Monte Carlo results show that 
not only space but also time emerges dynamically
in this model.
Furthermore, 
the real-time dynamics extracted from the matrices
turns out to be remarkable:
3 out of 9 spatial directions start to expand at some critical time.
This can be interpreted as the birth of our Universe.
%

\keywords{Superstring theory; matrix model; nonperturbative effects.}
\end{abstract}

\ccode{PACS numbers: 11.25.-w; 11.25.Sq}

\section{Introduction}

Fundamental questions concerning our Universe are:
\begin{romanlist}[(ii)]
\item Why does our Universe have (3+1)-dimensions?
\item Why is it expanding ?
\end{romanlist}
We provide explicit answers to these questions
from a nonperturbative formulation of superstring theory
in (9+1)-dimensions \cite{Kim:2011cr}.

More specifically, we use
the type IIB matrix model \cite{IKKT}, 
which was proposed as a 
nonperturbative formulation of string theory based on
type IIB superstrings in 10 dimensions.
The Euclidean version of this model has been studied 
by various approaches. 
In particular, the Gaussian expansion method proposed in 
Ref.~\refcite{Nishimura:2001sx} 
was used to calculate
the free energy 
of the SO($d$) symmetric vacua ($2 \le d \le 7$)
and it was found that $d=3$ gives 
the minimum \cite{Nishimura:2011xy}.
This implies
that the SO(10) symmetry of the Euclidean model
is spontaneously broken down to SO(3).
Moreover, the extent of space-time in the extended $d$ directions
and that in the shrunken $(10-d)$ directions 
turn out to have a finite ratio
even in the large-$N$ limit \cite{Nishimura:2011xy}.
While these results reveal interesting dynamical 
properties of the Euclidean
model, the connection to our real space-time is not very clear.
This motivated us to consider the Lorentzian model \cite{Kim:2011cr}.

Let us quote here an interesting statement made by Seiberg
in his rapporteur talk entitled 
``Emergent Spacetime'' \cite{Seiberg:2006wf}
at the 23rd Solvay Conference in Physics in 2005:
\emph{Understanding how time emerges will undoubtedly shed new light
on some of the most important questions in
theoretical physics including the origin of the Universe.}
Indeed, we will see that
the emergent spacetime is naturally realized in the
Lorentzian matrix model, and that the results can be 
interpreted as describing the birth of our Universe.

The rest of this article is organized as follows.
In section \ref{sec:prev-works}
we review previous works on type IIB matrix model.
In section \ref{sec:def-Lorentzian} we define the Lorentzian matrix model.
In section \ref{sec:Monte} we present our Monte Carlo results
for the Lorentzian matrix model.
Section \ref{sec:summary} 
is devoted to a summary and discussions.

\section{Previous works on type IIB matrix model}
\label{sec:prev-works}


The action of the type IIB matrix model is given by \cite{IKKT} 
%
\beqa
S &=& S_{\rm b} + S_{\rm f} \ , 
\label{action}
 \\
S_{\rm b} &=& -\frac{1}{4} \, \tr \Bigl( [A_{\mu},A_{\nu}]
[A^{\mu},A^{\nu}] \Bigr) \ , 
\label{b-action}
\\
S_{\rm f}  &=& - \frac{1}{2} \,
\tr \Bigl( \Psi _\alpha (\, {\cal C} \,  \Gamma^{\mu})_{\alpha\beta}
[A_{\mu},\Psi _\beta] \Bigr)  \ ,
\label{f-action}
\eeqa
where  $A_\mu$ ($\mu = 0,\cdots, 9$) and
$\Psi_\alpha$ ($\alpha = 1,\cdots , 16$) are
$N \times N$ 
Hermitian matrices.
The Lorentz indices $\mu$ and $\nu$ are
contracted
using the metric
$\eta={\rm diag}(-1 , 1 , \cdots , 1)$.
The $16 \times 16$ matrices $\Gamma ^\mu$ are
ten-dimensional gamma matrices after the Weyl projection,
and the unitary matrix ${\cal C}$ is the
charge conjugation matrix.
The action has manifest SO(9,1) symmetry,
where $A_{\mu}$ and $\Psi _\alpha$ transform as a
vector and
a Majorana-Weyl spinor, respectively.

There are various evidences
that the model gives a nonperturbative formulation of 
superstring theory \cite{IKKT}.\footnote{There are also other proposals
for nonperturbative formulations of superstring/M theory
based on matrix models such as Matrix Theory \cite{BFSS}
and Matrix String Theory \cite{Dijkgraaf:1997vv}.}
First of all, the action (\ref{action})
can be viewed as a matrix regularization of the worldsheet action
\beq
S_{\rm Schild} = - \int d^2 \xi \sqrt{g}
\Big( \frac{1}{4} \{ X_\mu , X_\nu \}   \{ X^\mu ,X^\nu \} 
+ \frac{1}{2} \Psi {\cal C} \Gamma^\mu  \{ X_\mu , \Psi \} \Bigr) 
\label{Schild}
\eeq
of type IIB superstring theory in a particular gauge known 
as the Schild gauge. 
%
(The ``Poisson brackets'' are defined here by
$\{ f(\xi) , g(\xi) \} \equiv 
\epsilon_{ij} \frac{\partial f}{\partial \xi_i} 
\frac{\partial g}{\partial \xi_j}$.)
It has also been argued that
configurations of block-diagonal matrices
correspond to a collection of disconnected worldsheets 
with arbitrary genus.
Therefore, instead of being 
equivalent
just to the worldsheet theory,
the large-$N$ limit of the matrix model is expected to be 
a second-quantized theory of type IIB superstrings,
which includes multi-string states.
Secondly, D-branes are represented as classical solutions in the 
matrix model, and the interaction between them calculated
at one loop reproduced correctly the known results
from type IIB superstring theory \cite{IKKT}.
Thirdly, one can derive
the light-cone string field theory for the type IIB case
from the matrix model \cite{Fukuma:1997en} with a few assumptions.
In the matrix model, one can define
the Wilson loops, which can be naturally identified
with the creation and annihilation operators of strings.
Then, from the Schwinger-Dyson equations for the Wilson loops,
one can actually obtain the string field Hamiltonian. 

In all these connections to string theory,
it is crucial that the model has two kinds of
fermionic symmetries given by
\begin{align}
& \left\{ \begin{array}{ll}
\delta^{(1)}A_{\mu} &=i\epsilon_1 {\cal C}\Gamma_{\mu} \Psi  \ , \\
\delta^{(1)}\Psi &=\frac{i}{2}\Gamma^{\mu\nu}[A_{\mu},A_{\nu}]\epsilon_1 \ , 
\end{array} \right.
\label{susy1} \\
& \left\{ \begin{array}{ll}
\delta^{(2)}A_{\mu} &=0 \ , \\
\delta^{(2)}\Psi &=\epsilon_2 \1 \ ,
\end{array} \right.
\label{susy2}
\end{align}
where $\1$ is the unit matrix.
It also has the bosonic symmetry given by
\begin{align}
\left\{ \begin{array}{ll}
\delta^{(3)}A_{\mu} &=c_{\mu} \1  \ , \\
\delta^{(3)}\Psi &=0 \ .
\end{array} \right.
\label{translation}
\end{align}
Let us denote the generators of (\ref{susy1}), 
(\ref{susy2}) and (\ref{translation}) by 
$Q^{(1)}$, $Q^{(2)}$ and $P_{\mu}$, respectively, 
and define their linear combinations 
\begin{align}
\tilde{Q}^{(1)}=Q^{(1)}+Q^{(2)} \ , 
\;\;\; 
\tilde{Q}^{(2)}=i(Q^{(1)}-Q^{(2)}) \ .
\end{align}
Then, we find that the generators satisfy the algebra
\begin{align}
[\epsilon_1{\cal C}\tilde{Q}^{(i)},\epsilon_2{\cal C}\tilde{Q}^{(j)}] =
-2\delta^{ij}\epsilon_1{\cal C}\Gamma^{\mu}\epsilon_2P_{\mu} \ ,
\label{N2SUSY}
\end{align}
where $i,\, j=1,2$.
This is nothing but the ten-dimensional ${\cal N}=2$ supersymmetry.
It is known that field theories with
this symmetry
necessarily include gravity, which suggests that
so does the type IIB matrix model.
When we identify (\ref{N2SUSY}) with 
the ten-dimensional ${\cal N}=2$ supersymmetry,
the symmetry (\ref{translation}) is identified
with the translational symmetry in ten dimensions,
which implies that 
the eigenvalues of $A_{\mu}$ should 
be identified with 
the coordinates of ten-dimensional space-time.
This identification 
is consistent with the one adopted in stating
the evidences listed in the previous paragraph,
and shall be used throughout this article as well.

An interesting feature of the type IIB matrix model
is that the space-time itself is treated as a part of
dynamical degrees of freedom in the matrices.
Therefore, it is possible that a four-dimensional
space-time is generated dynamically
in this model.
This issue has been studied in the Euclidean version of 
the model \cite{AIKKT,Hotta:1998en,Ambjorn:2000bf,%
Ambjorn:2000dx,Anagnostopoulos:2001yb,Anagnostopoulos:2010ux,%
Anagnostopoulos:2011cn,%
Nishimura:2000ds,Nishimura:2000wf,Nishimura:2001sq,%
Nishimura:2001sx,Kawai:2002jk,Aoyama:2006rk,%
Imai:2003vr,Imai:2003jb,Imai:2003ja,Nishimura:2011xy},
which can be obtained from (\ref{action})
by making a Wick rotation
\beq
A_0 \mapsto iA_{10} \ , \quad \quad \Gamma^0 \mapsto -i\Gamma^{10} \ .
\eeq
Note that the Euclidean model has SO(10) symmetry instead of 
SO(9,1).

In order to discuss the spontaneous symmetry breaking (SSB)
of $\SO(10)$ in the large-$N$ limit, 
we consider the 
``moment of inertia'' tensor \cite{AIKKT,Hotta:1998en}
\begin{equation}
  T_{\mu\nu} = \frac{1}{N} \tr (A_\mu A_\nu) \ , 
\label{eq:tmunu}
\end{equation}
which is a $10\times10$ real symmetric tensor. 
We denote its eigenvalues as $\lambda_j$ ($j=1,\cdots, 10$) 
with the specific order 
\begin{equation}
  \lambda_1 \geq \lambda_2 \geq \cdots \geq \lambda_{10}  \ .
\label{eq:lambda}
\end{equation}
If the SO(10) symmetry is not spontaneously broken, 
the expectation values $\langle \lambda_j  \rangle$ 
($j=1, \cdots , 10$) should be all equal in the large-$N$ limit.
The dynamical generation of $d$-dimensional space-time corresponds to
an SO($d$) symmetric vacuum,
in which the expectation values 
$\langle \lambda_j  \rangle$ ($j=1, \cdots , d$) 
are equal in the large-$N$ limit,
but the remaining ones $\langle \lambda_j  \rangle$ 
($j=d+1, \cdots , 10$) are much smaller.

\begin{figure}[tb]
\begin{center}
\begin{tabular}{cc}
\epsfig{file=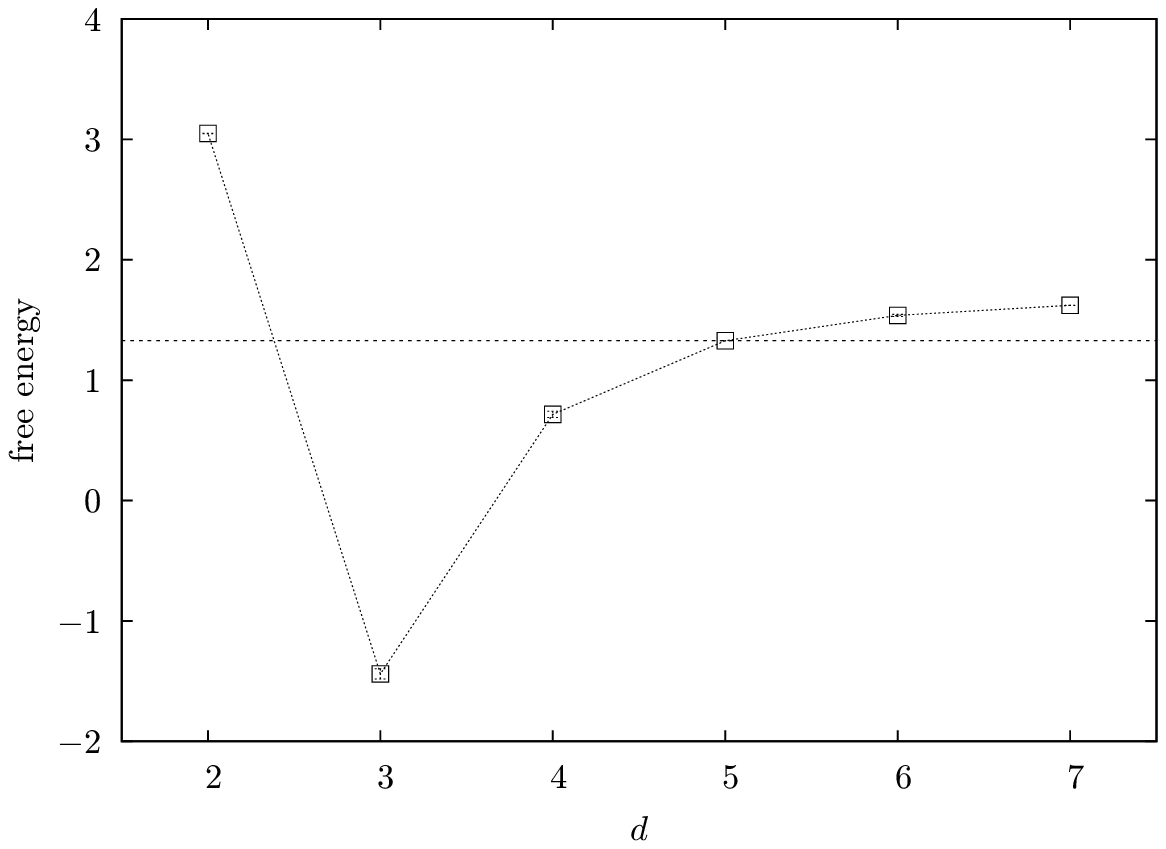,width=.47\textwidth} &
\epsfig{file=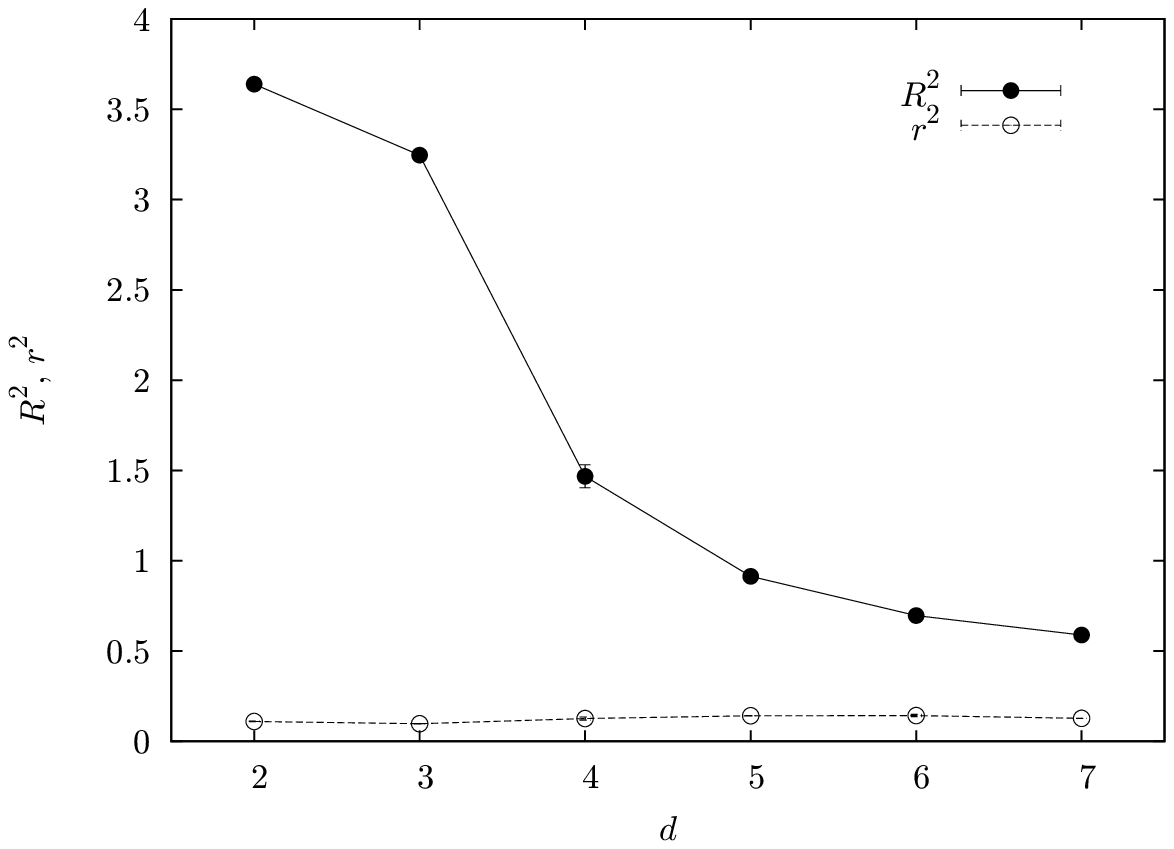,width=.47\textwidth} \\
\end{tabular}
\end{center}
\caption{
(Left) The free energy 
of the SO($d$) symmetric vacuum
is plotted against $d$. 
The horizontal
line represents the value $f=\log 8 - \frac{3}{4} = 1.32944\ldots$
obtained from the conjecture 
by Krauth, Nicolai and Staudacher \cite{Krauth:1998xh}.
(See footnote \ref{KNS-conjecture}.)
The dotted line connecting
the data points is drawn to guide the eye.
(Right) The extent of space-time $R^2$ and $r^2$ 
in the extended
and shrunken directions, respectively, 
are plotted against $d$.
The solid and dashed lines connecting
the data points are drawn to guide the eye.
}
\label{fig:free-summary}
\end{figure}

Let us show recent results obtained by the Gaussian expansion 
method \cite{Nishimura:2011xy}. 
In Fig.~\ref{fig:free-summary} (Left) we plot
the free energy of the SO($d$) symmetric vacuum 
for $2 \le d \le 7$
at order 3 of the expansion.
The result decreases monotonically as $d$ decreases from 7 to 3,
and it becomes much larger for $d=2$.
Thus, the $\SO(3)$ symmetric vacuum gives the smallest free energy,
which suggests the SSB of
SO(10) down to SO(3).\footnote{\label{KNS-conjecture}The $d$-dependence of the free energy
is quite analogous to the one observed in the six-dimensional
model \cite{Aoyama:2010ry}.
There the value of the free energy tends to decrease
slightly as one goes from order 3 to order 5.
Considering such artifacts due to truncation,
we speculate that 
the Krauth-Nicolai-Staudacher conjecture \cite{Krauth:1998xh}
actually 
refers to the partition function for the ${\rm SO}(10)$ symmetric vacuum.
See Fig.~\ref{fig:free-summary} (Left).}

Let us discuss the results for the extent of space-time represented
by the eigenvalues (\ref{eq:lambda}).
In Fig.~\ref{fig:free-summary} (Right)
we plot the result for the extended directions ($R^2$) and 
the shrunken directions ($r^2$) for each $d$. 
We find that $r^2$ stays almost constant at $r^2 = 0.1 \sim 0.15$,
which seems to be universal for all the SO($d$) symmetric
vacua with $2 \le d \le 7$.
On the other hand, the results for $R^2$
are found to
be larger for smaller $d$.
Thus the extent of space-time seems to be finite in all directions.
While the observed SSB of SO(10) is an interesting
dynamical property of the Euclidean model, 
the connection to the real space-time is unclear.

\section{Defining the Lorentzian matrix model}
\label{sec:def-Lorentzian}

The most crucial difference between the
Euclidean and Lorentzian models is
the bosonic part of the action (\ref{b-action}).
In the Euclidean model, it becomes positive semi-definite
\beq
S_{\rm b} \propto  \tr ( F_{\mu\nu} )^2  \ ,
\eeq
where we have defined Hermitian matrices 
$F_{\mu\nu} = i [A_\mu , A_\nu]$.
There is a flat direction corresponding to $[A_\mu , A_\nu]\sim 0$,
but it is lifted up by quantum effects 
from fermionic zero modes \cite{AIKKT}.
Thus the model is well defined without any 
cutoff \cite{Krauth:1998xh,AW}.

In the Lorentzian model, 
the bosonic action can be decomposed into two terms
\beq
S_{\rm b} 
\propto  \tr ( F_{\mu\nu} F^{\mu\nu} )
= - 2 \, \tr ( F_{0i} )^2 + 
 \tr ( F_{ij} )^2 
\label{bosonic-action}
\eeq
with opposite signs.
The model looks extremely unstable, and hence
no one has ever dared to study this model seriously.

Another important difference between the Euclidean 
and Lorentzian models is the Pfaffian
${\rm Pf}{\cal M}(A)$
obtained by integrating out fermions.
In the Euclidean model, the Pfaffian is complex,
and the phase plays a crucial role in SSB of 
SO(10) \cite{Nishimura:2000ds,Nishimura:2000wf,Nishimura:2001sq}.
But it also makes Monte Carlo studies extremely 
difficult \cite{Anagnostopoulos:2001yb,%
Anagnostopoulos:2010ux,Anagnostopoulos:2011cn}
due to the so-called sign problem.
In the Lorentzian model, the Pfaffian is real.
This is a good news for Monte Carlo studies, but we
also lose a source of SSB.

%
In the case of Euclidean model, 
the partition function was defined by
\beq
Z_{\rm E} = \int d A \, d\Psi \, e^{- S} =
\int d A \,  e^{- S_{\rm b}} {\rm Pf}{\cal M}(A) \ .
\label{Euclidean-partition-fn-def}
\eeq
In the case of Lorentzian model,
we define the partition function of the Lorentzian model 
by \cite{Kim:2011cr}
\beq
Z_{\rm L} = \int d A \, d\Psi \, e^{i S} =
\int d A \,  e^{i S_{\rm b}} {\rm Pf}{\cal M}(A) \ .
\label{partition-fn-def}
\eeq
Here 
we have replaced the ``Boltzmann weight'' $e^{-S}$
used in the Euclidean model by $e^{iS}$.
This is theoretically motivated
from the connection
to the worldsheet theory (\ref{Schild}).
%
When we make an inverse Wick rotation,
we need to change the worldsheet coordinate $\xi_0 \equiv - i\xi_2$
as well as the target-space coordinates.
Applying the mapping rule from fields to matrices \cite{IKKT},
we obtain (\ref{partition-fn-def}).

Note that $e^{iS_{\rm b}}$ in the partition function (\ref{partition-fn-def})
is a phase factor
just as in the path-integral formulation of quantum field theories
in Minkowski space.
This may give rise to the sign problem
when one tries to study the model by Monte Carlo simulation.
In the present case, however, the sign problem 
can actually be circumvented in the following way.
The crucial point here 
is that
the action of the type IIB matrix model
is homogeneous
with respect to the matrices.
Under the scale transformation $A_\mu \mapsto \rho A_\mu$,
each part of the partition function (\ref{partition-fn-def}) 
transforms as
\beqa
S_{\rm b} &\mapsto&\rho ^4 S_{\rm b}  \ ,  \\
dA &\mapsto&\rho ^{10(N^2-1)} dA  \ , \\
{\rm Pf}{\cal M}(A) &\mapsto&\rho ^{8(N^2-1)} {\rm Pf}{\cal M}(A) \ .
\eeqa
Integrating out the scale factor
of the bosonic matrices first,\footnote{The same
procedure was also used in Ref.~\refcite{Krauth:1998xh} 
for simulating the Euclidean model.}
one essentially converts
the phase factor $e^{i S_{\rm b}}$ into a constraint
$S_{\rm b} \approx 0$.
(Such a constraint is analogous
to the one that appeared in the model inspired by space-time uncertainty
principle \cite{Yoneya:1997gs}.)

It turns out that the integration over $A_\mu$ in
(\ref{partition-fn-def}) is divergent,
and we need to introduce two constraints
\beqa
\frac{1}{N}\tr (A_0)^2  &\le&  \kappa \frac{1}{N} \tr (A_i)^2  \ ,
\label{T-constr} \\
\frac{1}{N} \tr (A_i)^2   &\le&  L^2  \ .
\label{R-constr}
\eeqa
This is in striking contrast to the Euclidean model,
in which the partition function is shown to
be finite without any regularization \cite{Krauth:1998xh,AW}.

Without loss of generality, we set $L=1$ in (\ref{R-constr}), 
and thus we arrive at the model
\beq
Z =  \int dA \,
\delta \left(
\frac{1}{N}\tr (F_{\mu\nu}F^{\mu\nu})  \right)
 {\rm Pf} {\cal M} (A)
\, \delta\left(\frac{1}{N}\tr (A_i)^2 - 1 \right)
\theta\left(\kappa  - \frac{1}{N}\tr (A_0)^2  \right)  \ ,
\label{our-model}
\eeq
where $\theta(x)$ is the Heaviside step function.
Since the Pfaffian ${\rm Pf} {\cal M}(A)$ is real
in the present Lorentzian case,
the model (\ref{our-model}) can be studied by Monte Carlo simulation
without the sign problem.\footnote{Strictly speaking, the Pfaffian
can flip its sign, but we find that the configurations with
positive Pfaffian dominates as $N$ is increased.
Hence, we just take the absolute value of the Pfaffian in actual simulation.} 
Note that this is usually not the case for quantum field theories in
Minkowski space.

\section{Monte Carlo results for the Lorentzian matrix model}
\label{sec:Monte}

We perform Monte Carlo simulation\footnote{For a recent review
on various applications of Monte Carlo simulation to string theory,
see Ref.~\refcite{Nishimura:2012xs}.}
of the model (\ref{our-model})
by using the 
RHMC algorithm \cite{Clark:2003na}.

It turned out that not only space but also time emerges dynamically
in the Lorentzian matrix model.
We found that the eigenvalue distribution 
of $A_0$ extends in the large-$N$ limit.
Here, supersymmetry of the model plays a crucial role.
If we omit fermions, the eigenvalue distribution has 
a finite extent, and the cutoff (\ref{T-constr})
in the temporal direction is actually not needed \cite{KNT}.

\begin{figure}[tb]
\begin{center}
\begin{tabular}{cc}
\epsfig{file=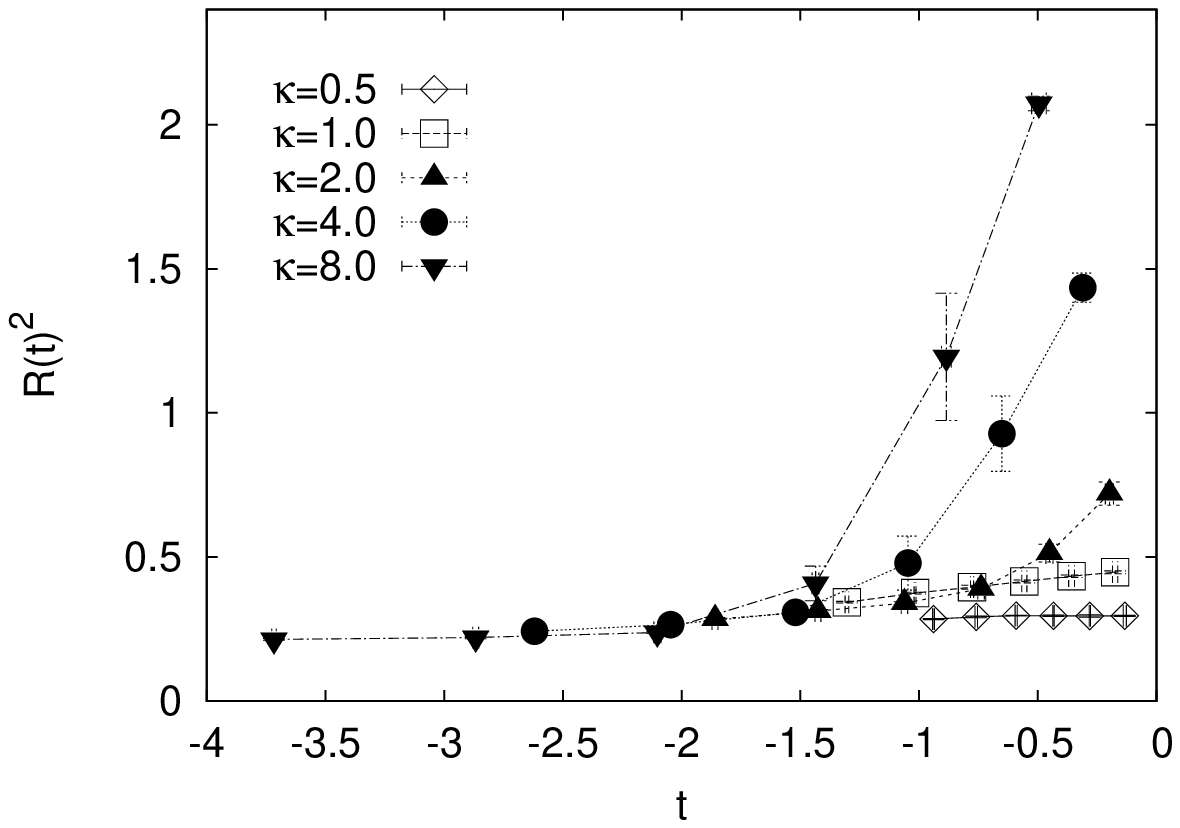,width=.47\textwidth} &
\epsfig{file=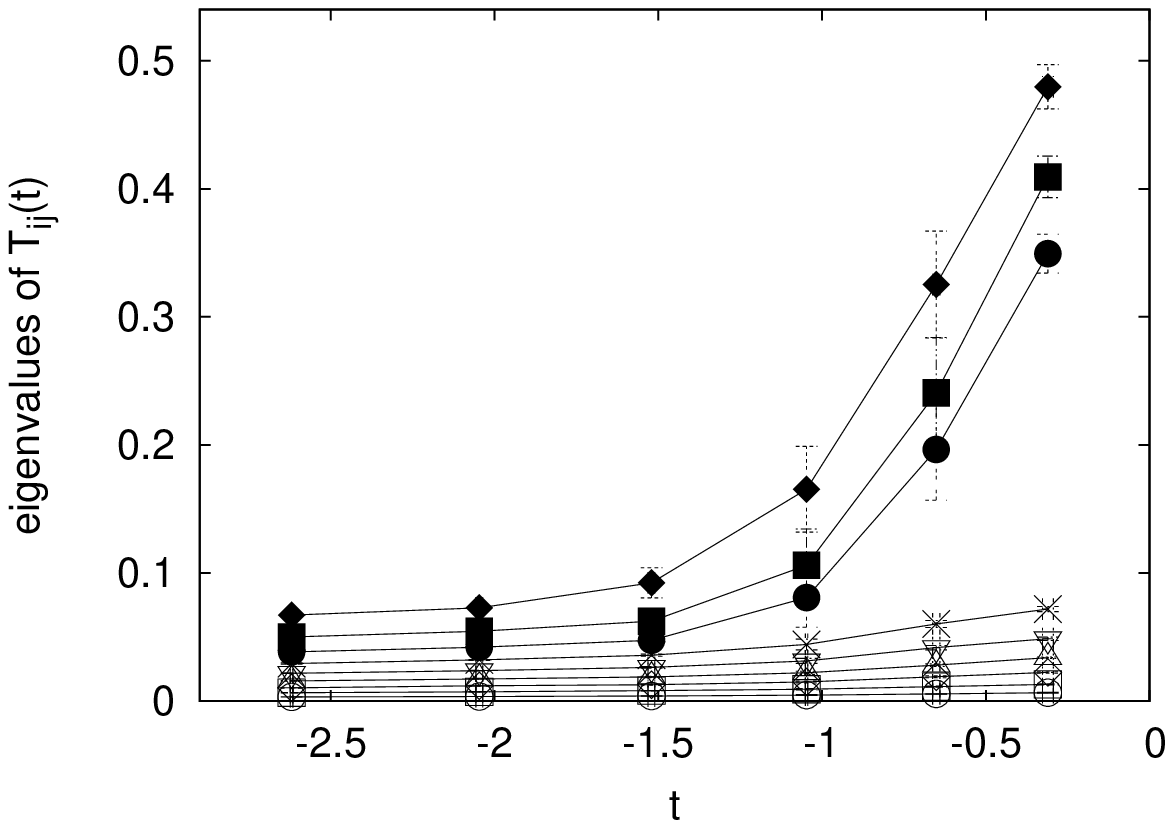,width=.47\textwidth} \\
\end{tabular}
\end{center}
\caption{
(Left) The extent of space $R(t)^2$ with $N=16$ and $n=4$
is plotted as a function of $t$
for five values of $\kappa$.
The peak at $t=0$ starts to grow at some critical $\kappa$.
(Right) The 9 eigenvalues of $T_{ij}(t)$ with $N=16$ and $n=4$
are plotted as a function of $t$
for $\kappa=4.0$. After the critical time $t_{\rm c}$,
3 eigenvalues become larger,
suggesting that the SO(9) symmetry is spontaneously broken
down to SO(3).
}
\label{Rt}
\end{figure}

In order to extract the ``time evolution'', we diagonalize $A_0$,
and define the eigenvectors $| t_a \rangle$ corresponding
to the eigenvalues $t_a$ of $A_0$ ($a=1 , \cdots , N$)
with the specific order $t_1 < \cdots < t_N$.
The spatial matrix in this basis $\langle t_{a} | A_i | t_{b} \rangle $
is not diagonal, but it turns out that the off-diagonal elements
decrease rapidly as one goes away from a diagonal element.
This motivates us to
define $n\times n$ matrices
\beq
\bar{A}_i^{(ab)}(t) \equiv  \langle t_{\nu+a} | A_i | t_{\nu+b} \rangle 
\eeq
with $1 \le a , b \le n$ and
$t= \frac{1}{n}\sum_{a=1}^{n} t_{\nu + a}$
for $\nu=0,\cdots , (N-n)$.
These matrices represent the 9d space structure 
at fixed time $t$.
(This point of view can be justified in the large-$N$ limit,
in which more and more eigenvalues of $A_0$ 
appear around some value $t$
within a fixed interval $\delta t$.)
The block size $n$ should be large enough to include non-negligible 
off-diagonal elements.
In Fig.~\ref{Rt} (Left) we plot the extent of space 
\beq
R(t)^2 \equiv \frac{1}{n} \tr \bar{A}_i(t)^2
\eeq
for $N=16$ and $n=4$.
Since the result is symmetric under the time reflection
$t \rightarrow -t$ as a consequence of the symmetry $A_0 \rightarrow -A_0$,
we only show the results for $t<0$.
There is a critical $\kappa$, beyond which
the peak at $t=0$ starts to grow.

Next we study the SSB of the SO(9) symmetry.
As an order parameter,
we define the $9 \times 9$
(positive semi-definite)
real symmetric tensor
\beq
T_{ij}(t) = \frac{1}{n} \tr \Bigl\{
\bar{A}_i(t) \bar{A}_j(t) \Bigr\} \ .
\eeq
The 9 eigenvalues of $T_{ij}(t)$ are
plotted against $t$ in Fig.~\ref{Rt} (Right) for $\kappa=4.0$.
We find that 3 largest eigenvalues of
$T_{ij}(t)$ start to grow at the critical
time $t_{\rm c}$, which suggests that the SO(9)
symmetry is spontaneously broken down to
SO(3) after $t_{\rm c}$.
Note that $R(t)^2$ is given by the sum of 9 eigenvalues
of $T_{ij}(t)$.

\begin{figure}[tb]
\begin{center}
\begin{tabular}{cc}
\epsfig{file=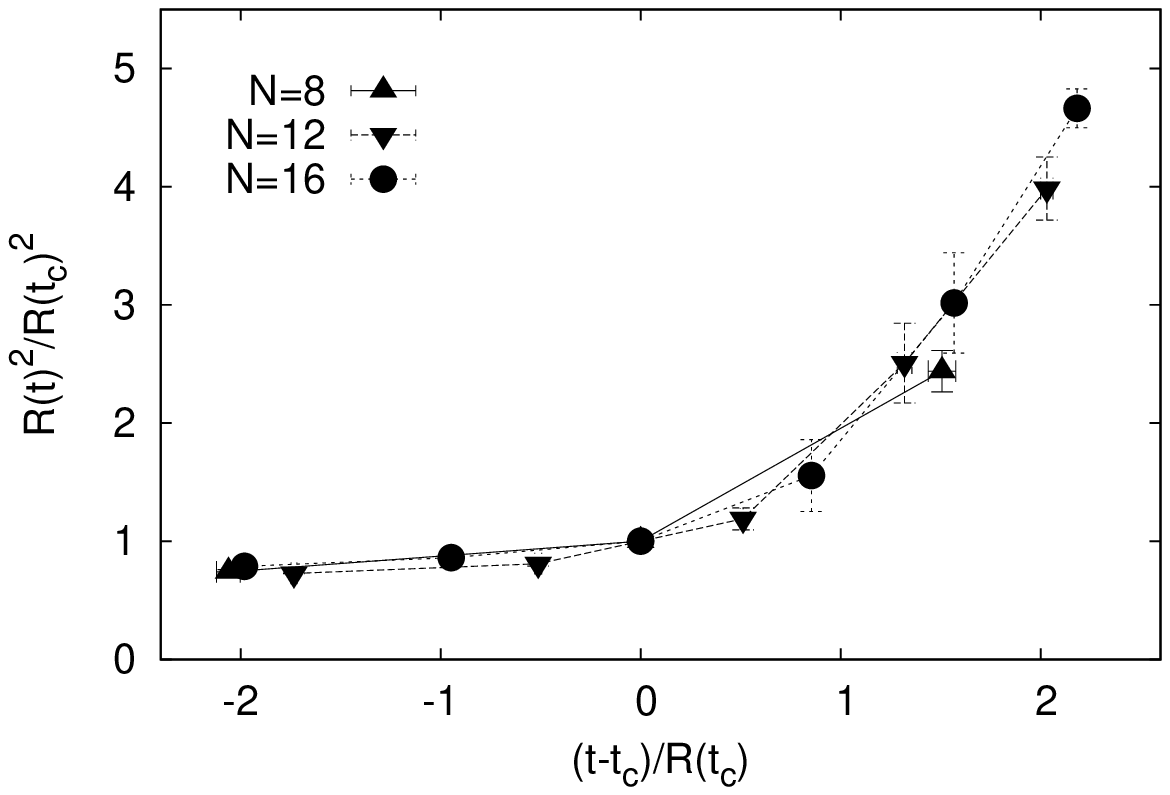,width=.47\textwidth} &
\epsfig{file=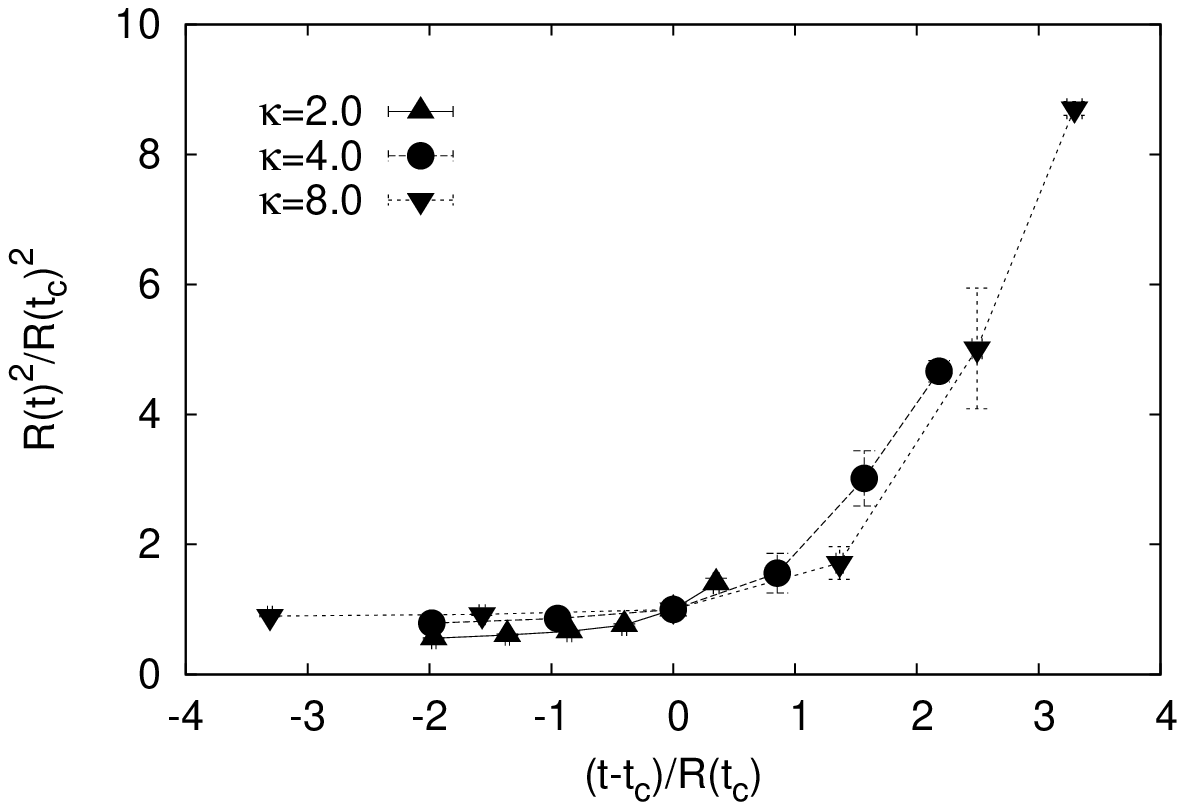,width=.47\textwidth} \\
\end{tabular}
\end{center}
\caption{
(Left) The extent of space $R(t)^2$ 
for $\kappa=\beta N^{1/4}$ is plotted for $N=8,12,16$ with $\beta=2$.
We plot the results against the
shifted time $t - t_{\rm c}$ in units of
the size of the universe $R(t_{\rm c})$
at the critical time.
(Right) Similar plot for fixed $N=16$ with $\kappa=2,4,8$.
}
\label{Rt-rescaled}
\end{figure}

It turned out that one can remove the
infrared cutoffs $\kappa$ and $L$ 
in the large-$N$ limit
in such a way that $R(t)$ scales.
This can be done in two steps.
(i) First we send $\kappa$ to $\infty$ with $N$ as
$\kappa = \beta \, N^{p}$ ($p\simeq \frac{1}{4}$) \cite{KNT}.
The scaling behavior is clearly seen in Fig.~\ref{Rt-rescaled} (Left).
The scaling curve of $R(t)$ one obtains in this way
depends on $\beta$.
(ii) Next we send $\beta$ to $\infty$ with $L$.
The two limits correspond to
the continuum limit and
the infinite volume limit, respectively, in quantum field theory.
Thus the two constraints (\ref{T-constr}), (\ref{R-constr})
can be removed in the large-$N$ limit,
and 
the resulting theory has no parameter other than one scale parameter.

Let us discuss the second limit (ii) in more detail.
We find that the inequality (\ref{R-constr}) is actually saturated 
for the dominant configurations.
Therefore, one only has to make the rescaling $A_\mu \mapsto L A_\mu$
in order to translate the configurations in the model (\ref{our-model})
as those in the original partition function.
It turns out that $R(t)$ for the rescaled configurations scales
in $\beta$ by tuning $L$ and shifting $t$ appropriately.
In order to see this, it is convenient to 
choose $L$ so that $R(t)$ 
at the critical time $t=t_{\rm c}$ becomes unity,
and to shift $t$ so that the critical time comes to the origin.
Then $R(t)$ with increasing $\beta$ 
extends in $t$
in such a way that the results at smaller $|t|$ scale.
This is demonstrated in Fig.~\ref{Rt-rescaled} (Right),
where we find a reasonable scaling behavior
for $N=16$ with $\kappa=2.0, 4.0, 8.0$.
Note, in particular, that the extent of ``time'' increases
as $\kappa$ is increased,
which is not the case 
in the bosonic model \cite{KNT}.
As we mentioned earlier, 
supersymmetry plays a crucial role for the ``emergent time'' 
in the Lorentzian matrix model.


The SSB of SO(9) looks mysterious at first sight, but
we can actually understand the mechanism quite intuitively.
Let us consider the case in which $\kappa$ is large.
Then the first term of (\ref{bosonic-action}) becomes
a large negative value, and therefore the second term
has to become large in order to 
make (\ref{bosonic-action}) zero as required in (\ref{our-model}).
Due to the constraint $\frac{1}{N} \tr (A_i)^2=1$, however,
it is more efficient to maximize the second term of (\ref{bosonic-action})
at some fixed time.
The system actually chooses the middle point $t=0$, where
the suppression on $A_i$ from the first term of (\ref{bosonic-action})
becomes the least.
This explains why the peak of $R(t)$ at $t=0$ grows as we increase $\kappa$.

Let us then consider a simplified question: what is the configuration of $A_i$
which gives the maximum $\frac{1}{N} \tr (F_{ij})^2$ with fixed
$\frac{1}{N} \tr (A_i)^2=1$. Using the Lagrange multiplier $\lambda$,
we maximize the function $G=  \tr (F_{ij})^2 - \lambda \,  \tr (A_i)^2$.
Taking the derivative with respect to $A_i$, we obtain
$2\, [A_j,[A_j, A_i]] - \lambda A_i = 0 $.
This equation can be solved if $A_i = \chi L_i$ for $i\le d$,
and $A_i = 0$ for $d < i \le 9$,
where $L_i$ are the representation matrices
of a compact semi-simple Lie algebra with $d$ generators.
Clearly $d$ should be less than or equal to 9.
It turns out that the maximum of
$\frac{1}{N} \tr (F_{ij})^2$ is achieved for the SU(2) algebra,
which has $d=3$,
with $L_i$ being the direct sum of the spin-$\frac{1}{2}$ representation
and $(N-2)$ copies of the trivial representation.
This implies the SSB of SO(9) down to SO(3).
The SSB
can thus be understood
as a classical effect in the $\kappa\rightarrow \infty$ limit.
%
When we tune $\kappa$ with increasing $N$ as described above,
quantum effects become important.
We have confirmed \cite{KNT} that the $n\times n$ matrix
$Q = \sum_{i=1}^9 \bar{A}_i(t)^2$ has 
quite a continuous eigenvalue distribution,
which implies that the space is not like a two-dimensional sphere as
one might
suspect from the classical picture.




\section{Summary}
\label{sec:summary}

We have studied the type IIB matrix model
as a nonperturbative formulation of 
superstring theory in (9+1)-dimensions.
Unlike previous works, we made the Lorentzian matrix model
well-defined by introducing infrared cutoffs instead of making
a Wick rotation.

Monte Carlo studies of the Lorentzian matrix model
revealed the following nontrivial facts:
\begin{itemlist}
 \item The infrared cutoffs can be removed in the large-$N$ limit.
 \item The theory thus obtained has no dimensionless parameter,
supporting the validity of the model as a nonperturbative formulation of
superstring theory.
\item ``Time'' emerges dynamically thanks to supersymmetry.
\item ``Time evolution'' of 9d space emerges.
\item 3 out of 9 spatial directions start to expand at some 
``critical time'', after which the SO(9) symmetry of the space
is broken spontaneously down to SO(3).
\end{itemlist}

All these results for the Lorentzian matrix model
suggest that (3+1)-dimensional expanding
universe emerges dynamically from superstring
theory if the theory is treated nonperturbatively.
This may be contrasted with the quantum cosmology 
in the early 80s \cite{Vilenkin:1982de,Vilenkin:1984wp,Hartle:1983ai}
that aimed at describing the birth of the universe
within the mini-superspace approximation.\footnote{More recently, 
a nonperturbative approach to quantum gravity
has been pursued
using the causal dynamical triangulation \cite{Ambjorn:2005qt}.
For earlier works that put forward the idea to use matrices
for cosmology, 
see Refs.~\refcite{Freedman:2004xg,Craps:2005wd}.
See also Refs.~\refcite{Steinacker:2010rh,Lee:2010zf} 
for related works on emergent gravity.}
Note also that 
the picture suggested here
is quite different from that 
in (perturbative) superstring theory, where space-time
with various dimensions can be obtained 
by compactification or by using D-brane backgrounds.

The rapid expansion of the three-dimensional space
observed in Monte Carlo simulation 
may be 
interpreted
as the beginning of inflation.
It would be interesting to investigate
the microscopic origin of the inflation 
along this line. 

The mechanism of the SSB
relies crucially on 
noncommutativity of 
the space-time represented by 10 bosonic matrices.
Therefore, 
an important issue
is whether the usual commutative space-time appears at later times.
We addressed this issue by studying the classical equations
of motion, which
are expected to be valid
at late times \cite{Kim:2011ts,Kim:2012mw}.
There are actually infinitely many solutions representing
commutative (3+1)-dimensional space-time.
Moreover, we found a simple solution 
with an expanding behavior, which
naturally solves the cosmological constant problem \cite{Kim:2012mw}.
We consider that there exists a unique solution of this kind
that dominates
the partition function of the matrix model
at late times.
%
By pursuing this direction further, 
it would be possible to understand the origin of
dark energy found in the present cosmological observations
and to predict the fate of our Universe.

%
Since superstring theory is not only a theory of quantum gravity
but also a theory of all the matters and the fundamental 
interactions among them,
it would be interesting to see how the Standard Model
appears at late times in
the Lorentzian matrix model.
Finding solutions to the classical equation of motions
\cite{Aoki:2010gv,Steinacker:2011wb,Chatzistavrakidis:2011gs,%
Chatzistavrakidis:2011su,Aoki:2012ei}
and performing perturbative expansion around them
would be an important direction as an approach complementary to
Monte Carlo simulation.

Recently, we have proposed an explicit
procedure to identify the local fields corresponding to the 
massless modes that appear at late times \cite{Nishimura:2012rs}.
The basic assumption is that the low-lying spectrum is essentially
determined by the Nambu-Goldstone modes (and their extension) 
associated with the 
SSB of the (9+1)-dimensional Poincare symmetry
and supersymmetry.
The local field theory obtained in this way below the Planck scale
has interesting generic features.
It is a grand unified theory with the gauge group SU($k$),
and all the matter fields are in the adjoint representation.
The grand unified theory with $k=8$, for instance, can accommodate
all the Standard Model particles.
As the space-time dimensionality seems to be uniquely determined by
nonperturbative dynamics of superstring theory,
we consider it quite conceivable that
the Standard Model emerges uniquely from this top-down approach.
Further investigations 
are clearly worth while.


To conclude, we believe that
the Lorentzian matrix model, as a correct and tractable
nonperturbative formulation of superstring theory,
provides 
totally new perspectives in both particle physics and cosmology.





\section*{Acknowledgments}
We would like to thank S.-W.\ Kim
and A.\ Tsuchiya for 
the fruitful collaborations that led to the main results 
discussed in this article.
This work is supported 
by Grant-in-Aid for Scientific
Research (No.\ 20540286 and 23244057) 
from Japan Society for the Promotion of Science.

\end{document}